\documentclass[referee]{aa}

\usepackage{times}
\usepackage{caption}
\usepackage[round]{natbib}
\usepackage{pdflscape}
\usepackage{afterpage}
\usepackage{capt-of}
\usepackage{color}
\usepackage{hyperref}
\usepackage{graphicx}
\usepackage{longtable}
\usepackage{tabu}
\usepackage{subfigure}
\usepackage{multirow}
\usepackage{amsmath}
\usepackage{verbatim}

\usepackage[normalem]{ulem}

\begin{document}

\title{Differentiation signatures in the Flora region}

\authorrunning{Oszkiewicz et al.}

\titlerunning{Differentiation signatures in the Flora region}

\author{Dagmara Oszkiewicz$^1$\and Pawe{\l} Kankiewicz$^2$\and Ireneusz W\l odarczyk$^3$\and \\ Agnieszka Kryszczy\'nska$^1$}

 \institute{Astronomical Observatory Institute, Faculty of Physics, A. Mickiewicz University, 
              S{\l}oneczna 36, 60-286 Pozna{\'n}, Poland
         \and Institute of Physics, Astrophysics Division, Jan Kochanowski University, Swietokrzyska 15, 25-406 Kielce, Poland
         \and Polish Astronomical Amateur Society, Powstancow Wlkp. 34, 63-708, Rozdra{\.z}ew, Poland
         }
         
\date{\today} 

 \offprints{D. Oszkiewicz, \\ e-mail: {\tt oszkiewi@astro.amu.edu.pl}}

   \date{Received xx xx xx / Accepted 28 09 2015} 
   
 \abstract
 {{\bf{ Most asteroid families are very homogeneous in physical properties. Some show greater diversity, however. The Flora family is the most intriguing of them. The Flora family is spread widely in the inner main belt, has a rich collisional history, and is one of the most taxonomically diverse regions in the main belt. As a result of its proximity to the asteroid (4) Vesta (the only currently known intact differentiated asteroid) and its family, migration between the two regions is possible.  This dynamical path is one of the counter arguments to the hypothesis that there may be traces of a differentiated parent body other than Vesta in the inner main belt region. We here investigate the possibility that some of the V- and A- types (commonly interpreted as basaltoids and dunites - parts of the mantle and crust of differentiated parent bodies) in the Flora dynamical region are not dynamically connected to Vesta.
 }}}
 {{\bf{The goal of this study is to investigate asteroids in the Flora dynamical region that may be witness to the differentiation of a parent body other than (4) Vesta. In particular, we aim at predicting which asteroids may be fragments of a differentiated body or bodies (taxonomical V-types). We also investigate their possible dynamical linkage to the nearby Vesta family.
 }}}
 {{\bf{To predict the taxonomic types of asteroids we used the naive Bayes classifier. We studied their dynamical past through numerical integration including gravitational and Yarkovsky forces. Each asteroid was cloned, and average orbital elements were computed at each step. When possible, we used observationally constrained physical parameters for the thermal forces.
 }}}
{{\bf{Most of the asteroids in the Flora region are predicted to originate from the S-complex (around 47.8\%). Tens of asteroids
with potentially differential origins are identified in the region, including 164 potential V types. We investigated the dynamical evolution of selected objects to examine possible linkages of selected asteroids with the nearby Vesta family. In addition to the candidate differentiated objects, we also studied the
dynamical evolution of those asteroids from the Flora family for which spectroscopic classification exist. In particular, we used all the physical parameters available for asteroid (809) Lundia to show that the asteroid is unlikely to originate in the Vesta region and that it resided on a stable orbit in the Flora dynamical region for at least 80 My. Its observationally constrained prograde rotation shows that asteroid (809) Lundia is unlikely to be a former Vesta family member.
}}}
{{\bf{We found that the V-type candidates are plentiful in the Flora dynamical region. Dynamical investigations of selected objects showed that some of these objects were likely not connected to asteroid (4) Vesta and its family for at least 100 My. This adds another piece of evidence to the hypothesis that there are
multiple differentiated parent bodies.}}}



\maketitle

\section{Introduction}

The Flora family is one of the asteroid families that were discovered early \citep{hirayama1918groups}. It is a remnant from a collisional disruption of a large body or bodies that occurred hundreds of My to a few Gy ago (the age of the family was approximated by several authors, for example $<1$Gy \citep{nesvorny2002flora} or $910^{+160}_{-120}$My \citep{dykhuis2014defining} ) .
The family occupies large area in the orbital element space and is the most dispersed asteroid family known \citep{nesvorny2002flora}. 
Because of its proximity to the $\nu_6$ resonance, it is commonly thought that the Flora region may be the source of many ordinary chondrites \citep{nesvorny2002flora} and near-Earth asteroids \citep{LaSpina2004, Kryszczynska2013}.  The dynamical and compositional links between the Flora asteroids, near-Earth asteroids, and ordinary chondrites led to multiple studies suggesting that the Flora family might be the source of many Earth impactors, such as the  Chelyabinsk meteorite \citep{reddy2013composition, reddy2014chelyabinsk} or a Cretaceous/Tertiary (K/T) impactor \citep{bottke2007asteroid, reddy2009composition}.

Early studies (in the 1970s through 1990s) of asteroids in the Flora region considered a possible differentiation scenario for a few asteroids in the Flora family. The final classifications showed no proofs for such a process, however.

In particular, several asteroids in the family were studied in detail
with spectral analyses. Of these, (951) Gaspra, an S-type asteroid, is the most thoroughly examined.
The Galileo spacecraft imaged the asteroid in four bands. Studies of these images showed a color variability of material on Gaspra's surface. 
\citet{carr1994geology} concluded that the variability is most likely due to variations in the thickness of space weathered regolith. 
A few other studies attributed the variability to a high olivine content and proposed a differential origin hypothesis \citep{veverka1994galileo, granahan1994galileo}.
This hypothesis was also supported by \citet{kivelson1993magnetic}, who showed that the magnetic properties of Gaspra as measured by the Galileo probe are consistent with stony iron meteorites.
The link between Gaspra and stony iron meteorites was refuted
by \citet{hood1994galileo}, however, who showed that the magnetic properties are also consistent with ordinary chondrites.

Another extensively studied asteroid from the Flora family is (8) Flora, also of S-type, the largest asteroid in the family. Similarly to (951) Gaspra, a possible differentiation of (8) Flora has also been widely discussed.
\cite{mccord1974asteroids} inferred Flora's composition to be a mixture of metal and pyroxine based on reflectance spectroscopy.
\cite{gaffey1984rotational} observed the spectral variability of (8) Flora with rotation and suggested that thermal alternations caused by a differentiation process shaped the mineralogy of the asteroid.

Many other asteroids in the Flora family have been identified as S-types. Some of them showed a spectrum of ordinary chondrites, such as (43) Ariadne  \citep{johnson1973spectrophotometry} (later
classified as Sk-type \citep{xu1995small, bus2002phase} and now considered an interloper \citep{hanuvs2013anisotropic}).
Later studies showed that the S-type asteroids can be divided into several subtypes. This spectral diversity was thought to originate from the exposure of materials of various degrees of differentiation \citep{gaffey1993mineralogical}. Adding to the complexity of the mineralogical interpretation of S-type asteroid spectra, \citet{chapman1996s} found that an S-type spectrum could result from a space-weathering mechanism acting on ordinary chondrite materials. More recently, multiple studies showed that impact shocking can also significantly effect the spectrum of ordinary chondrites \citep{kohout2014mineralogy}. As a result of the various mechanisms, the mineralogical identification of spectrally observed asteroids is challenging. Ordinary chondrites in particular are among the meteorite types whose spectra are the most affected by space weathering and impact shocking.

Despite the difficulties involved in interpreting the spectra, systematic spectral observations were undertaken to explain the origin of the Flora family.
Based on an analysis of UBV colors \citet{tedesco1979origin} concluded that the Flora population is mostly color homogenous. \citet{xu1995small} observed spectra of 20 asteroids from the Flora family. All of them except for (2259) Sofievka were found to belong to the S-complex. \citet{florczak1998visible} performed visible spectroscopy of 47 members of the Flora family. Most of the observed asteroids showed characteristics typical of S-complex objects. Five of the observed members displayed different spectral behavior. Of these, (4278) Harvey showed a V-type spectrum, (3533) Toyota showed an M-type spectrum, (3875) Staehle is of G or B type, (2093) Genichek is of C-type, and (298) Baptistina is of
E, M, or C type. \citet{florczak2002discovering} identified (809) Lundia and (956) Elisa as V-type asteroids in the Flora family and suggested that their origin might be in the Vesta family. \cite{mothe2005reanalysis} analyzed the taxonomic types present in asteroid families and clumps in the Flora family. Within the Baptistina family, \cite{mothe2005reanalysis} found objects of classes Xc, X, C, L, S, V, and A, making it one of the most taxonomically diverse families in the study.  Recently, \citet{dykhuis2014defining} redefined the core boundaries of the family using both orbital and reflectance properties. The results support a more spectrally homogenous definition of the family. 

On the other hand, meteoritic evidence shows that there might be objects other than those of the S complex in the region. Based on the Bunburra Rockhole meteorite fall observations and sample recovery, \citet{bland2009anomalous} suggested that they may be surviving fragments of a differentiated parent body or bodies in the inner main belt different from those that form (4) Vesta. 
Based on triangulated observations of the fall, \citet{bland2009anomalous} determined the pre-entry orbit and estimated a probability of
98\%\ that the observed meteoroid originated in the inner main belt and chances 72\% of it originating from the $\nu_6$ resonance (a main contributor to transporting asteroids from the Flora region \citep{nesvorny2002flora, scholl1991nu}). Furthermore, the meteorite oxygen isotopic composition in the recovered meteorites is significantly higher than that of howardite-eucrite-diogenite
(HED) meteorites, which are commonly interpreted to be part of the Vesta family. This leads to the conclusion that there must be V-type asteroids in the inner main belt that are not genetically related to (4) Vesta.


In this work we study the possibility that there might be traces of a differentiated parent body in the Flora dynamical region other than asteroid (4) Vesta. We focus on the dynamical Flora family members that were previously spectrally classified as V- and A-type asteroids as well as candidate V-type asteroids for the dynamical region. The V-type candidates were selected based on Sloan Digital Sky Survey multi-filter photometry, the
albedos were taken from the WISE, IRAS, and AKARI surveys and G$_{12}$ phase-curve parameters. Asteroids of V- and A-type are commonly interpreted to be basaltoids and dunites, parts of the crust and mantle of differentiated bodies (for example, \citet{mccord1970asteroid, bus2002phase, duffard2006v, lucas2012dunites}). Some ambiguity exists regarding the origin of A-type asteroids,
however. \citep{sanchez2014olivine} showed that the composition of A-type asteroids is consistent with brachinites and R chondrites, and \citep{sunshine2007olivine} showed that the R chondrites are formed by nebular processes. Therefore we mostly focus on V types.

To exclude a connection of the possible differentiated material to the Vesta family, we also analyze the dynamical history of these objects. In particular, we perform backward orbital propagation on a timescale of 100 My (typical family migration timescale).

In Sect. \ref{taxa} we describe the taxonomical analysis and select objects for dynamical investigation. In Sect. \ref{dyna} we investigate the dynamical paths of the asteroids and their past family ties. Conclusions are presented in Sect. \ref{con}.

\section{Taxonomic distribution}
\label{taxa}
\subsection{Data}

Throughout this study we use the dynamical definition of Flora family provided in \citet{Nesvorny2012} and definition of clumps and subfamilies provided in \cite{Diniz2012}. We decided to use these databases because they are only based on orbital elements to cluster asteroids into groups. This is a desired property because we search for traces of differentiation and therefore taxonomic variability. Asteroid families defined based on both dynamical and physical properties, such as \citet{masiero2013asteroid} may lead to a family classification that is taxonomically more homogeneous and may limit the possibilities of detecting traces of differentiation in dynamical regions. The dataset of \citet{Nesvorny2012}  contains 64 families calculated from analytic proper elements and 79 families calculated with synthetic elements  \citep{Nesvorny2012}. Asteroids are clustered using the hierarchical clustering method (HCM). The Flora cluster was computed based on analytic proper elements, and it contains around 16000 members, about 13000 of
which are numbered asteroids. We focus on the numbered asteroids because their orbits are well determined.

To taxonomically select and pre-classify V-type objects (detailed classification should be made when visible (VIS) and near-infrared (NIR) spectroscopy is available for these objects), we used albedos \citep{masiero2011main, usui2011asteroid, tedesco1994asteroid} and photometric measurements from the fourth release of the Moving Object Catalogue (MOC) of the Sloan Digital Sky Survey (SDSS) \citep{ivezic2001solar, abazajian2005third} and $G_{12}$ phase-curve parameters from \citet{oszkiewicz2011online}. 

We do not aim to derive an unbiased distribution of taxonomic types in the Flora family either by number or mass. Our main goal is to identify potential V-type objects in the Flora region and focus on investigating their dynamical connections (or lack of same) to the Vesta family. 

Because the inner main belt region is rich in dynamically overlaying families, there may be many interlopers present in the Flora family \citep{migliorini1995interlopers}. Therefore we would like to stress that the identified V-type asteroids may not be actual genetic members of the Flora family or any of the currently identified asteroid families. Our focus is to show that  some of the V types in this dynamical region did not migrate from (4) Vesta during their predictability period, not to identify their parent body or family.

\subsubsection{Geometric albedos}

\begin{figure}[htbp]
   \centering
   \includegraphics[width=0.5\textwidth]{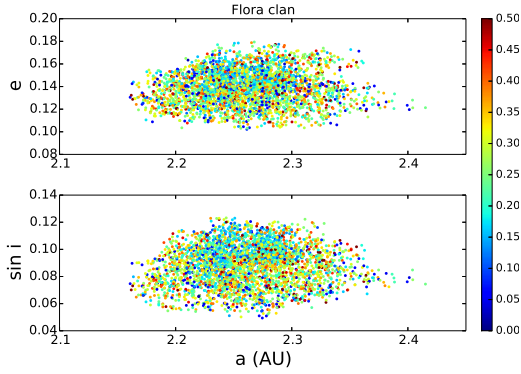} 
   \caption{Orbital element distribution of asteroids in the Flora region color-coded based on the albedo.}
   \label{albedo1}
\end{figure}

For all the numbered asteroids in the Flora family we extracted albedos from one of three databases: WISE \citep{masiero2011main}, AKARI \citep{usui2011asteroid}, and IRAS \citep{Tedesco2004} \citep{tedesco1994asteroid}. For each asteroid we extracted the albedo measurement with the smallest uncertainty from the three databases. This resulted in albedos for 3292 asteroids, 3202 of which are taken from WISE, 82 from AKARI, and 8 from IRAS. 
In general, using the different databases could lead to statistical biases in the resulting taxonomic distribution. However, we aim at predicting the taxonomic types for pre-selection purposes only and do not aim to derive an unbiased distribution. Therefore we did not perform consistency checks between the different databases. This approach is sufficient for the preselection purposes.
In Fig. \ref{albedo1} we present the distribution of proper orbital elements for asteroids in the Flora family, color-coded according
to albedo. The albedo values in the plots were restricted to range from $0$ to $0.5$ for image clarity. Very high albedos ($p_V\ge 0.5$) are listed for 55 asteroids (around $1.6$\% of the whole sample). The Baptistina family (median albedo values around $p_V=0.2$) clearly stands out in the albedo distribution plots. The median albedo of the whole sample in the Flora region is $p_V=0.25$, the standard deviation is $0.11$, and practically all albedo values are present in the sample. Median albedos for each of the subgroups and clumps in the Flora region for which we had at least $ten$ albedos are listed in Table \ref{averagepv}. For most of the clumps the standard deviation of  the albedo points is around 0.10. An exception is the Davidweilla clump, for which the albedo values are the lowest of all clumps, and both the average uncertainty ($0.03$) and the standard deviation of the sample ($0.05$) are small. Davidweilla is therefore more albedo homogenous than the other clumps in our sample. The Krinov clump also has a small standard deviation, but the average uncertainties are large.

\subsubsection{SDSS photometry}
We decided to use the most recent release of the SDSS MOC, which
is the fourth release. It contains data for $471,569$ moving objects, of which $220,101$ are linked to $104,449$ unique objects.
Following \cite{demeo2014solar} to guarantee good quality data, we removed points with the following flags:
\begin{verbatim}
EDGE BADSKY PEAKS_TOO_CLOSE NOTCHECKED 
BINNED4 NODEBLEND DEBLEND_DEGENERATE 
BAD_MOVING_FIT TOO_FEW_GOOD_DETECTIONS 
STATIONARY. 
\end{verbatim}
The flags meaning in the listed order: object was too close to edge of frame to be measured; should not affect point sources, local sky measurement failed, object photometry is meaningless,
some peaks were too close to be deblended, object contains pixels which were not checked for peaks by deblender; deblending may be unreliable,
detected in 4x4 binned frame; few are genuine astrophysical objects, object is a blend, but was not deblended because it is: too close to an edge (EDGE already set), too large	(TOO LARGE), or a child overlaps an edge (EDGE will be set), some peaks had degenerate templates, motion inconsistent with straight line, not deblended as moving, even though detected, no good centroid found in enough bands for motion determination, a "moving" object's velocity is consistent with zero.

This resulted in around $5000$ observations of the Flora family members that were linked to around $2500$ unique objects.
Additionally, we required that the  parameters a* and i-z were within bounds of (-0.3, 0.4) and (-0.6, 0.2), respectively, and the errors in all filters (except for the u filter) were 10\% or better. The u-filter wavelengths are not covered by current taxonomies. This reduced the sample to around 4800 Flora family measurements for 2500 unique asteroids.

\begin{figure}[htbp]
   \centering
   \includegraphics[width=0.5\textwidth]{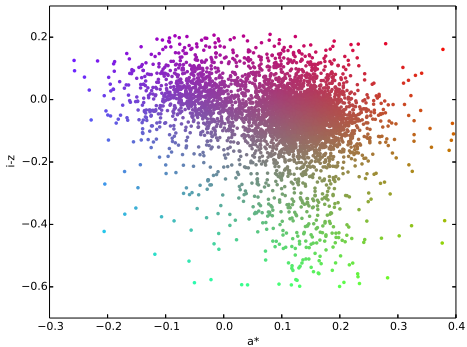} 
   \caption{Distribution of the a* (a* = 0.89 (g - r) + 0.45 (r - i) - 0.57 \cite{ivezic2001solar}) and i-z color index in the Flora family.}
   \label{SDSS1}
\end{figure}

In Table \ref{averagea} we list the median a* and i-z for asteroids in the Flora family and its subgroups. 
In Fig. \ref{SDSS1} we present the a* versus i-z color index distribution for the Flora family. Asteroids of all ranges of a* and i-z are present in the family. The vast majority of asteroids lies in the S-complex cloud.


\subsubsection{Phase curves}

Phase-curve parameters are available for around 500 000 asteroids \citep{oszkiewicz2011online}, of which around 11 000 are numbered Flora family members.
We required that the G$_{12}$ parameter is within $0.0$, to $1.0$ bounds and that both the right- and left-sided G$_{12}$ errors are smaller than 20\%. This reduced the sample to around 450 G$_{12}$ parameters (G$_{12}$s). Generally, all values of G$_{12}$s are present in the sample. The weighted average (weights are one over the sum of the right- and left-sided error) of the family is $\overline{G_{12}}=0.66,$ and the standard deviation is $0.10$.

\subsection{Predicted taxonomic distribution}

We classified all asteroids in the Flora family using the method described in \cite{oszkiewicz2014}. That is, we classified asteroids based on the available physical features such as albedo, reflectance values, and phase-curve parameters. The classification was made using the naive Bayes classifier. In particular, the taxonomic class was predicted by computing probabilities for all the Bus-DeMeo classes and selecting the most likely class for an object \citep{oszkiewicz2014}:
\begin{equation}
\text{arg max} \Bigg{(} P(C = c_k) \prod _{j=1}^n P(F_j | C = c_k) \Bigg{)}_{ \text{ for } k = 1,..,m}.
\end{equation}
where $ P(C = c_k)$ is the a priori probability that the object is of $c_k$ type, $P(F_j | C = c_k)$ is the conditional probability of the object having some specific value of a feature $F_j$ given that it belongs to type $c_k$, $n$ is the number of features, and $m$ is the number of types. We mostly used the SDSS photometry; when possible,  we added the geometric albedo $p_v$ and the phase-curve parameter G$_{12}$ to the set of features. There are about 150 objects for which a full set of features is available. Where several sets of observations were available for an object, we chose the classification with the highest posterior probability.

About 2500 Flora family members in total were assigned a predicted taxonomic type. The most common complex of the predicted types is S (1195 asteroids). The
preponderance of S-complex asteroids in the Flora family (including the Baptistina family) is expected because it corresponds to earlier studies. The remaining main complexes include X-complex (187 asteroids) and C-complex (152 asteroids). Asteroids from V- (164 asteroids), K- (78 asteroids), L- (393 asteroids), D- (155 asteroids), and A- (178 asteroids) types were also predicted. In Fig. \ref{types} we present the overall contributions of S-, X-, and C-complexes as well as end-member types. 

The high (6.2\%) percentage of D-type asteroids is a surprise at this distance from Sun. D-type objects are usually found far from the Sun and are generally associated with Trans-Neptunian objects (TNOs). \cite{demeo2014unexpected} followed up D-type candidates in the inner main belt predicted in \cite{demeo2013taxonomic} and found that only approximately 20\% can be observationally confirmed as true D types. We suspect that many of our predicted D types are also false positives.  Similarly, low confirmation rates may be found for other taxonomic types, especially those with no distinct features in the SDSS wavelength ranges. For example, predicted A-type asteroids are often found to be S-complex objects after the addition of NIR
wavelength data \citep{lucas2012dunites}. As a result of the distinct spectral characteristics in the visible wavelengths, the confirmation rates for the SDSS-predicted V types are much higher (for example, the survey by \cite{solontoi2012avast} achieved
a confirmation rate of 77\%) and therefore are more reliable. 

\begin{figure}[htbp]
   \centering
   \includegraphics[width=0.5\textwidth]{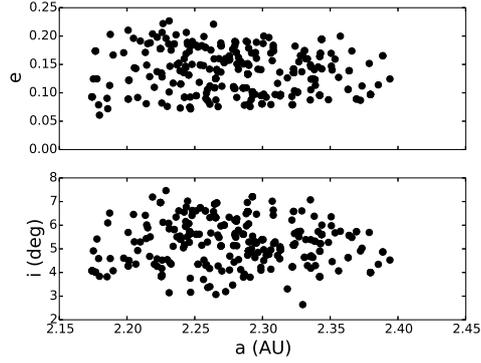} 
   \caption{Distribution of the predicted taxonomic V types in the Flora family.}
   \label{dist}
\end{figure}

In our study the predicted V types are 6.6\% of the total predicted types. This is quite a high percentage, and if spectrally confirmed (and if migration from Vesta can be dynamically excluded - we
discuss this in the next section), these objects could substantially strengthen the hypothesis of traces of a differentiated body other than Vesta in the inner main belt. In Fig. \ref{dist} we present the orbital element distribution for the predicted V types in the Flora region. Similarly to other types, the predicted V-type asteroids seem to be randomly distributed in the orbital element space. 

We compared the predicted number of V types in the Flora family with the fraction of modeled fugitives. \citet{nesvorny2008fugitives} modeled the evolution of  Vesta family members  over 1 Gy and showed that the modeled fraction of Vesta fugitives in the inner main belt region I (defined as 2.2$<$a$<$2.3 AU 0.05$<$e$<$0.2, 0$^{\circ}$$<$i$<$10$^{\circ}$) is $f_M=2.3\%$. This region approximately corresponds to half the Flora-family region in the semi-major axis. To estimate the uncertainties of the fraction, \citet{nesvorny2008fugitives} shifted the region in orbital element space. This procedure showed
a fugitive fraction of $f_M = 1.7\% - 3.1 \%$ . To compare our predicted number of V types with that of the fugitives predicted by \citep{nesvorny2008fugitives}, we computed the fraction of predicted V types in region I to the number of all predicted types in the region. The fraction is 7.44\% - more than twice the highest modeled fraction in the area. \citet{nesvorny2008fugitives} also found that the fugitives that evolve into the region I typically have high negative values of $da/dt$ Yarkovsky drift in the semi-major axis. If the V types in this region migrated from Vesta, they should have preferentially retrograde spins and near-optimal Yarkovsky drift parameters (i.e., an obliquity $\epsilon \approx 180^{\circ}$ and favorable thermal conductivity) to reach their current position in 1-2 Gy \citep{nesvorny2008fugitives}. In Sect \ref{dyna} we discuss the dynamical evolution of selected asteroids considering observationally constrained parameters (including obliquity) for modeling of the thermal forces.

\begin{figure}[htbp]
   \centering
   \includegraphics[width=0.5\textwidth]{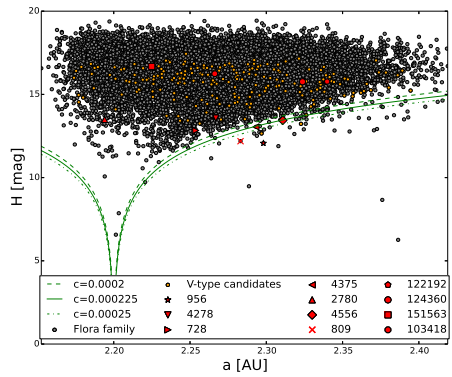} 
   \caption{Dependence of absolute magnitude H on the proper semi-major axis a for the Flora family.}
   \label{Ha}
\end{figure}

\begin{figure}[htbp]
   \centering
   \includegraphics[width=0.5\textwidth]{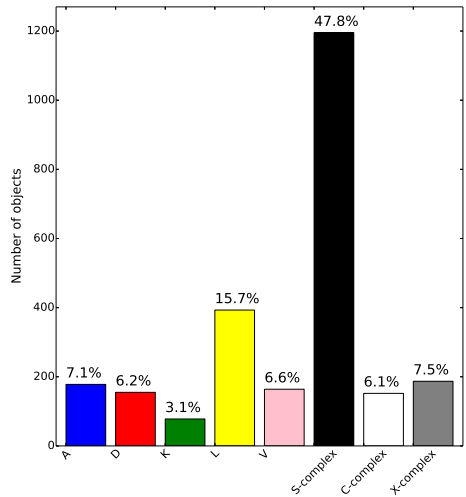} 
   \caption{Predicted taxonomic types in the Flora family.}
   \label{types}
\end{figure}

In Fig. \ref{Ha} we show the dependence of absolute magnitude and proper semi-major axis for the Flora family. We plot there
three different family borders. For different values of the C parameter relating to a different initial extent of the family or different ages and magnitudes of the Yarkovsky drift, we refer
to \cite{brovz2013constraining} . The left side of the "V-shape" plot for the Flora family is truncated due to the nearby $\nu_6$ resonance. Two asteroids, (809) Lundia and (956) Elisa, which previously have been confirmed as V types, lie outside the family borders. They both are large objects and therefore unlikely to be driven to the edges of the family by the Yarkovsky force. The two asteroids are even more distant from the borders of the Vesta family. Even though they have been identified as Flora family members using the hierarchical clustering method, they are likely to be interlopers in Flora and Vesta. 

With the exception of two objects, most of the V-type candidates lie within the Flora family borders. However, as a result of the complex dynamical nature of this region, these objects may not be actual genetical Flora family members. We study their dynamical history in Sect. \ref{dyna}.

For completeness we list asteroids in the Flora family in Table \ref{other}  with a spectrally determined taxonomic type. We do not list the most prevalent S- and C-complex asteroids in the table. C-complex objects are primitive undifferentiated objects, and S-complex asteroids may indicate partial differentiation. We here mostly focus on V types, which show evidence of full planetary differentiation.

\begin{figure}[htbp]
   \centering
   \includegraphics[width=0.5\textwidth]{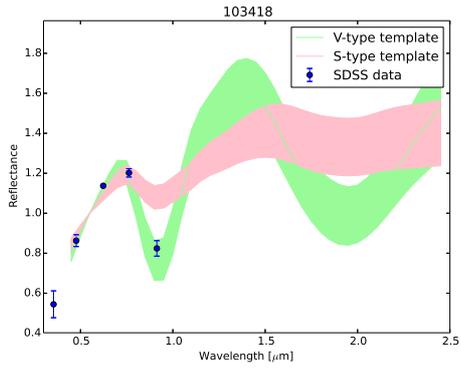} 
   \caption{V-type template fit to SDSS data of asteroid 103418.}
   \label{f1}
\end{figure}

\begin{figure}[htbp]
   \centering
   \includegraphics[width=0.5\textwidth]{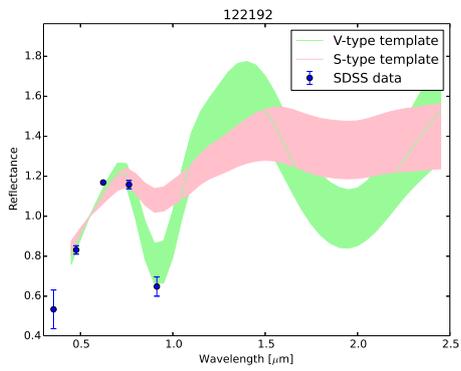} 
    \caption{V-type template fit to v data of asteroid 122192.}
   \label{f3}
\end{figure}

\begin{figure}[htbp]
   \centering
   \includegraphics[width=0.5\textwidth]{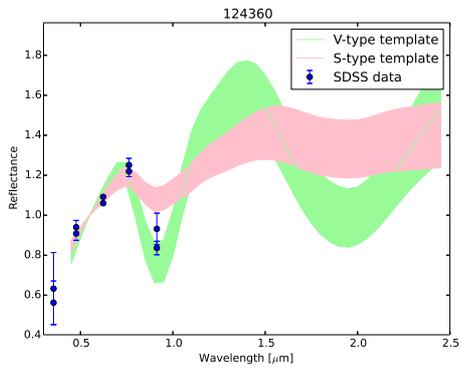} 
   \caption{V-type template fit to SDSS data of asteroid 124360.}
   \label{f4}
\end{figure}

\begin{figure}[htbp]
   \centering
   \includegraphics[width=0.5\textwidth]{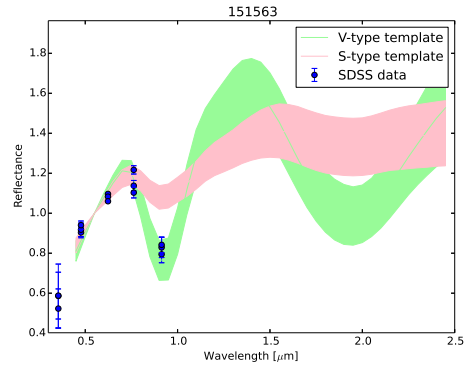} 
   \caption{V-type template fit to SDSS data of asteroid 151563.}
   \label{f5}
\end{figure}

For a more detailed investigation we selected 11 asteroids (listed in Tables \ref{table-elements} and \ref{table-thermal}). Although 164 V-type asteroids were predicted, a dynamical investigation can only be made for selected targets because the orbital propagation
is computationally
demanding  and time and resources are limited. We selected spectrally confirmed asteroids as well as V-type candidates indicated in this study. We tried to select objects for which some physical parameters are observationally constrained (albedo, size, or rotational period) for a better approximation of the thermal forces. Seven of the selected asteroids are of known taxonomic type (either V- or A-types), and four are of predicted V- ype (they were randomly selected from the pool of predicted V types). Fits to the V-type template for the predicted V types are presented in Figs. \ref{f1}-\ref{f5}, and they represent typical predicted V-type fits. The u-band points are plotted for reference only. To highlight the difference between the V  and S type (the most prevalent taxonomic type in this region), we have plotted the template for the S type as well. All of the selected asteroids had their next opposition in May to August 2015, predicted apparent V magnitudes between 17.5 - 19.5 mag, and can be targeted with large telescopes.

\section{Dynamical investigation}
\label{dyna}
The Flora region partially overlaps with the Vesta family in the orbital elements phase space. Therefore it has often been suggested that some of the asteroids currently associated with the Flora dynamical region could have originated in the Vesta family and then migrated to the Flora region. \cite{carruba2005v} presented a dynamical scenario in which the V-type asteroids (809) Lundia and (956) Elisa could have migrated from Vesta and therefore represent the same parent body. Other asteroids in the region, such as (4278) Harvey because of its V-type spectrum, were also suggested to originate from (4) Vesta \citep{florczak2002discovering}. In this section we test the migration scenario. In particular, we study the dynamical past of selected asteroids in the Flora region.

\subsection{Initial data}
We integrated the orbital elements of all the selected asteroids back in time up to 100 My into the past using two models: a simplified model (gravitational forces including the Sun and eight planets using the JPL DE405 planetary ephemeris) -- called 'grav. model', and a model with thermal forces (grav. model and Yarkovsky forces) -- called 'Yark. model'. The 100 My timescale is shorter than the latest estimates of Flora and Vesta ages (about 0.9 Gy for the Vesta family according to \cite{spoto2015asteroid}), but it is long enough to show the influence and direction of the Yarkovsky drift. We also took into account relatively low Lyapounov time ($L_T$) values of selected asteroids. The most 'predictable' is Lundia with $L_T$ = 552 ky, but for (151563) 2002 TM135, $L_T$ = 15.6 ky \citep{astdys}. Most of the selected asteroids have $L_T$ values on the order of $10^5$ years. Because
of these limitations, we decided to set the integration time to at most 100 My.

The starting orbital elements for the integration are listed in Table \ref{table-elements}. Initial orbital elements were computed using astrometric observations from the Minor Planet Center database (MPC) and the OrbFit software, version 4.2. \citep{Orbfit}. This new version includes the new error model based upon \citet{chesley2010treatment}. The starting orbital elements were computed using the JPL Planetary and Lunar Ephemerides DE405 and additional perturbations from the 25 massive asteroids. Their masses were taken from \citet{farnocchia2013yarkovsky}. 

We list in Table 5 the physical data of the asteroids that are
needed to reproduce perturbations from thermal forces: radius, bulk density, surface density, thermal conductivity, thermal capacity, albedo, infrared emissivity, period, and spin orientation. Where possible, we use modeled values for spin vectors and radius (otherwise random values were assumed). Densities and thermal properties were adopted from literature as 'typical' for particular taxonomic group. We made several assumptions. Asteroid shapes were assumed spherical for simplicity. The double asteroid (809) Lundia \citep{kryszczynska2009new, birlan2013spectroscopy} was approximated by a single body of an equivalent diameter of 9.1 km. 
This simplified model is sufficient for the current study. Our simplified model of (809) Lundia takes into account latest observational data (spin rate, rotational pole, diameter, and albedo), which
is an advantage compared with other studies. A more advanced simulation of the binary system including accurate shapes of both components and bYORP should be considered in the future. 

\subsection{Orbital integration}

The main tool used for the numerical integration was the swift\_rmvsy package \citep{Broz2011}. This tool is a modified and extended version of the swift\_rmvs that is included in the original SWIFT software \citep{Levison1994}. The initial step size of the swift\_rvmsy integrator was 20 days. This is a typical value in such calculations, especially for orbits of 
main belt asteroids. Details of Yarkovsky effect computations used in the swift\_rmvsy package were described by \cite{Vokrouhlicky1998} and \cite{Vokrouhlicky1999}.

For each asteroid we created 101 test particles (clones) following the procedure described by \cite{Milani2005}. The orbital distribution of clones was similar to real observational errors (Gaussian, with a 3-$\sigma$ criterion, along the line of variation). We analyzed the propagation of the mean values of the orbital elements. The elements were averaged and weighted using a Gaussian probability density function. For this reason, elements located close to the nominal solution were more important in weighting and averaging. 
We used averaged elements instead of proper elements because they are fast and easy to compute.
They also give a good overview and qualitative estimate of a
long-term dynamical evolution. Additionally, 
the separation of clones during the integration represents the propagation of observational errors.


\subsection{Dynamical evolution}
\begin{figure*}[b]
    \includegraphics[width=0.9\textwidth]{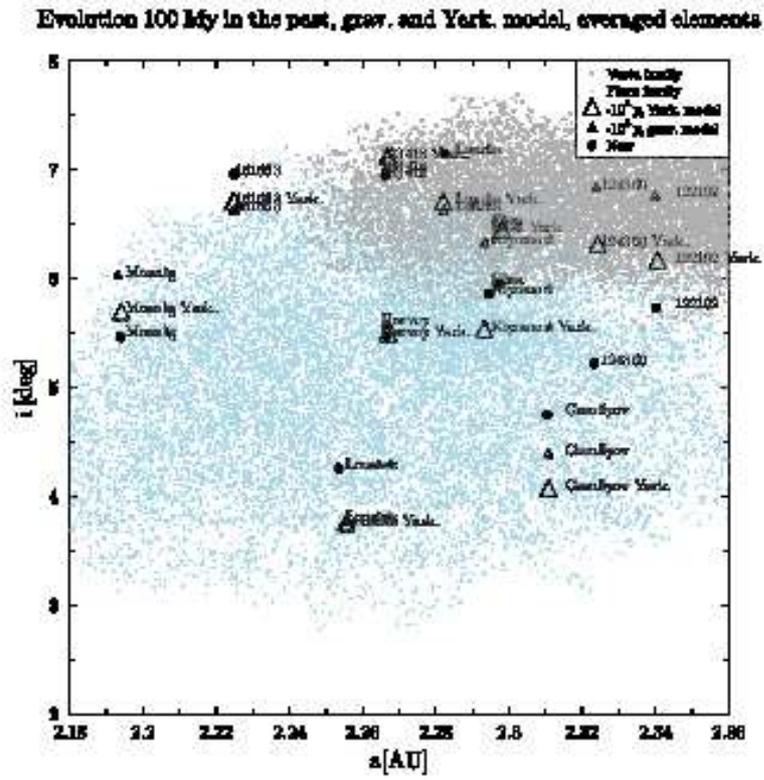}
    \caption{Orbital drift in semi-major axis $a$ and inclination $i$.}
    \label{all-orbits1}
\end{figure*}

\begin{figure*}[b]
    \includegraphics[width=0.9\textwidth]{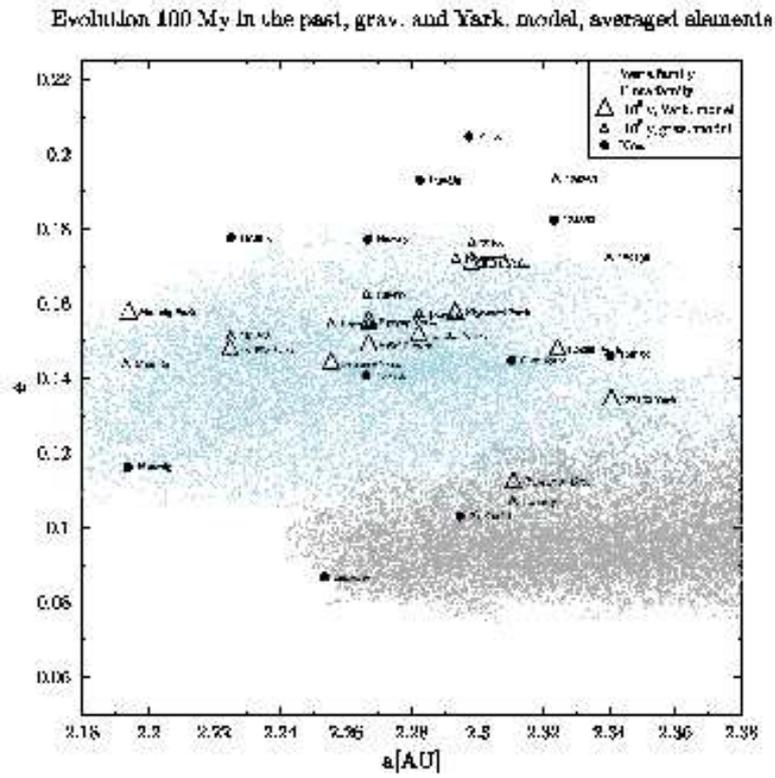}
    \caption{Orbital drift in semi-major axis $a$ and eccentricity $e$.}
    \label{all-orbits2}
\end{figure*}

In our sample of 11 asteroids and their clones, we did not find any collisions and ejections during the numerical integration. The numerical integration with the two dynamical models shows differences in the evolution of the averaged orbital elements. In Figs. \ref{all-orbits1} and \ref{all-orbits2} we present the results of the integration with the two models for all the 11 asteroids collectively.  Nine of these objects cluster very tightly in proper eccentricity $e$ and sine of inclination $\sin (i)$ near the Baptistina region (with the exception of 728 Leonisis and 4556 Gumilyov). For most of the asteroids, the Yarkovsky effect played an important role in the dynamical past. In Table \ref{drifts} we list the average values for the Yarkovsky drift in the semi-major axis for the 11 selected objects. We also show detailed results for three asteroids: (809) Lundia, (728) Leonisis, and (122192) 2000 KK82, as typical examples of the dynamical evolution in our sample.

\begin{figure*}[b]   
    \includegraphics[width=0.9\textwidth]{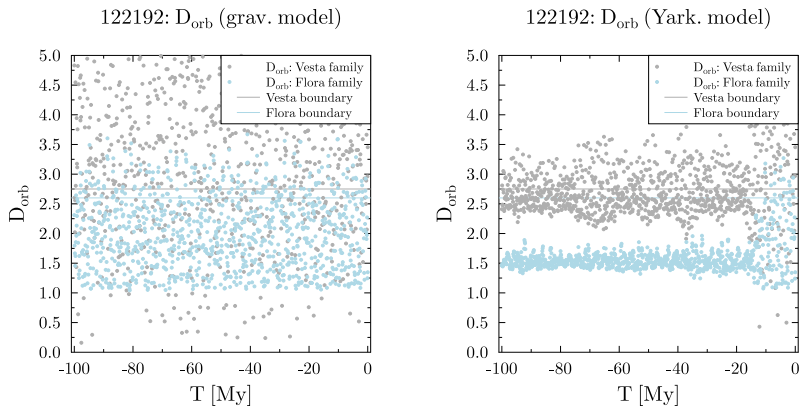}
    \caption{$D_{orbit}$ criteria of asteroid (122192) obtained with the grav. and Yark. models with regard to the Vesta and Flora families.\label{122192}} 
    \label{122192d}
\end{figure*}

\begin{figure*}[b]
    \includegraphics[width=0.9\textwidth]{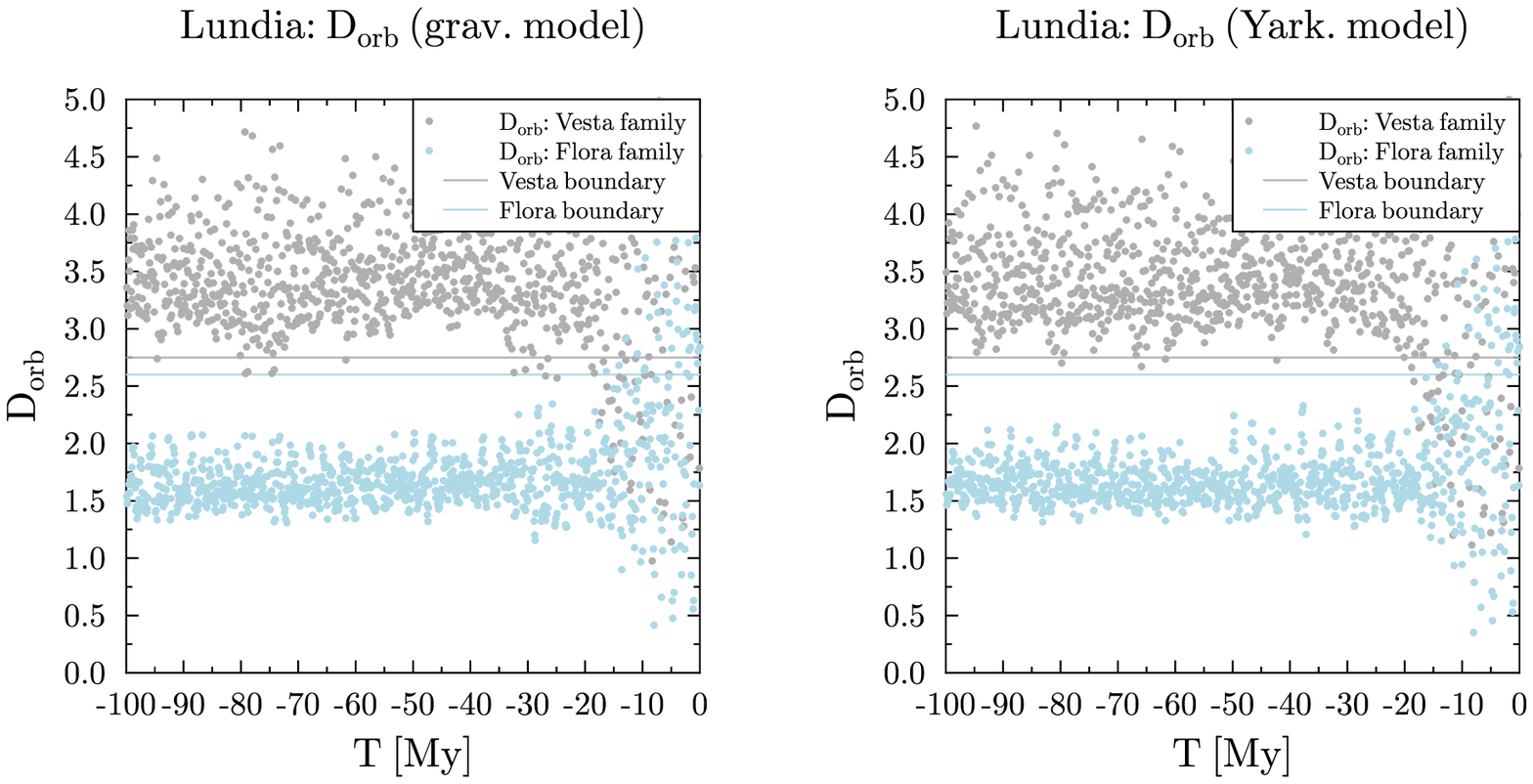}
    \caption{$D_{orbit}$ criteria of asteroid (809) Lundia obtained with the grav. and Yark. models with regard to the Vesta and Flora families.\label{lundia}} 
\end{figure*}

\begin{figure*}[b]
    \includegraphics[width=0.9\textwidth]{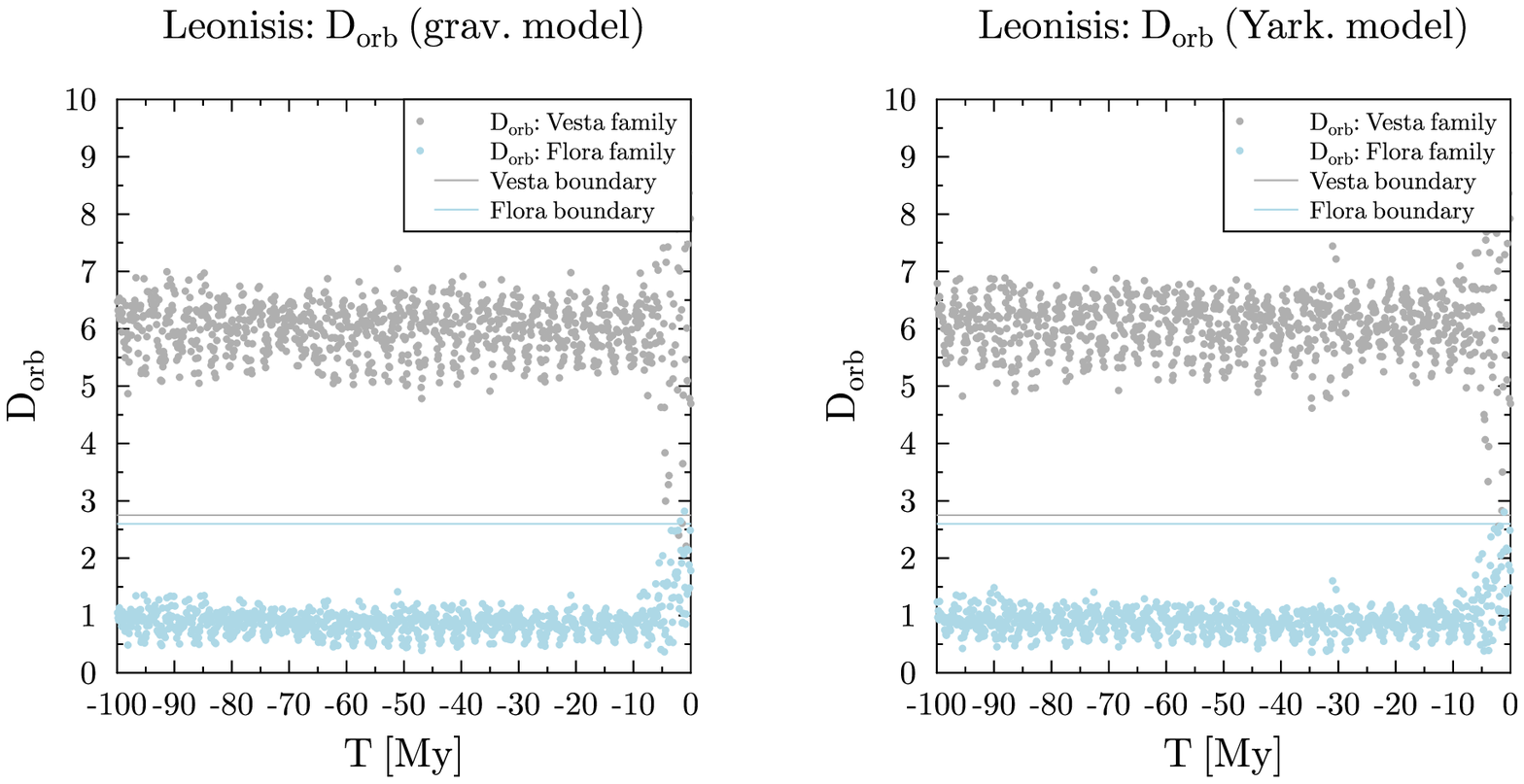}
    \caption{$D_{orbit}$ criteria of asteroid (728) Leonisis obtained with the grav. and Yark. models with regard to the Vesta and Flora families.\label{leonis}} 
\end{figure*}

\begin{figure*}[b]
    \centering       
    \includegraphics[width=0.4\textwidth]{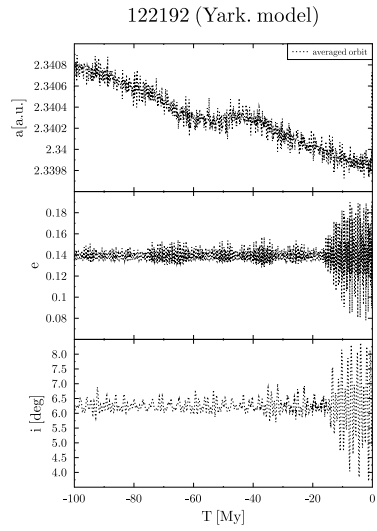}
\caption{Dynamic evolution of asteroid (122192) obtained with  the Yarkovsky model. \label{122192evolution}}
\end{figure*}

\begin{figure*}[b]
\centering       
    \includegraphics[width=0.4\textwidth]{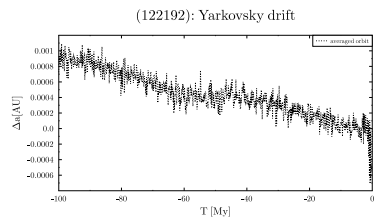}
    \caption{Drift in semi-major axis due to the Yarkovsky effect for asteroid (122192). \label{122192drift}} 
\end{figure*}

\begin{figure*}[b]
    \centering       
    \includegraphics[width=0.4\textwidth]{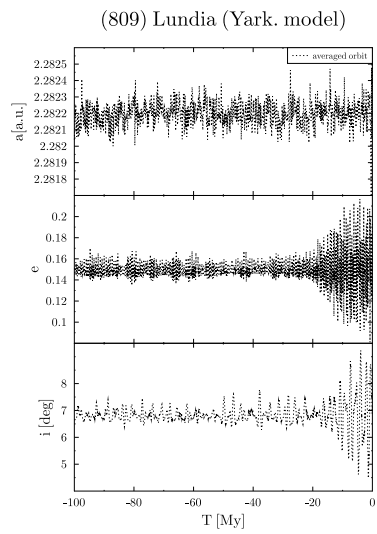}
\caption{Dynamic evolution of asteroid (809) Lundia obtained with  the Yarkovsky model. \label{809evolution}}
\end{figure*}

\begin{figure*}[b]
\centering       
    \includegraphics[width=0.4\textwidth]{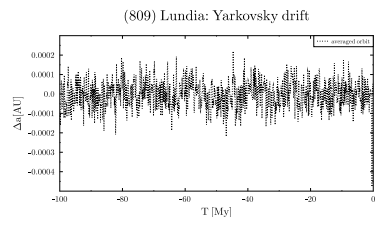}
    \caption{Drift in semi-major axis due to the Yarkovsky effect for asteroid (809) Lundia. \label{809drift}} 
\end{figure*}

\begin{figure*}[b]
\centering       
    \includegraphics[width=0.4\textwidth]{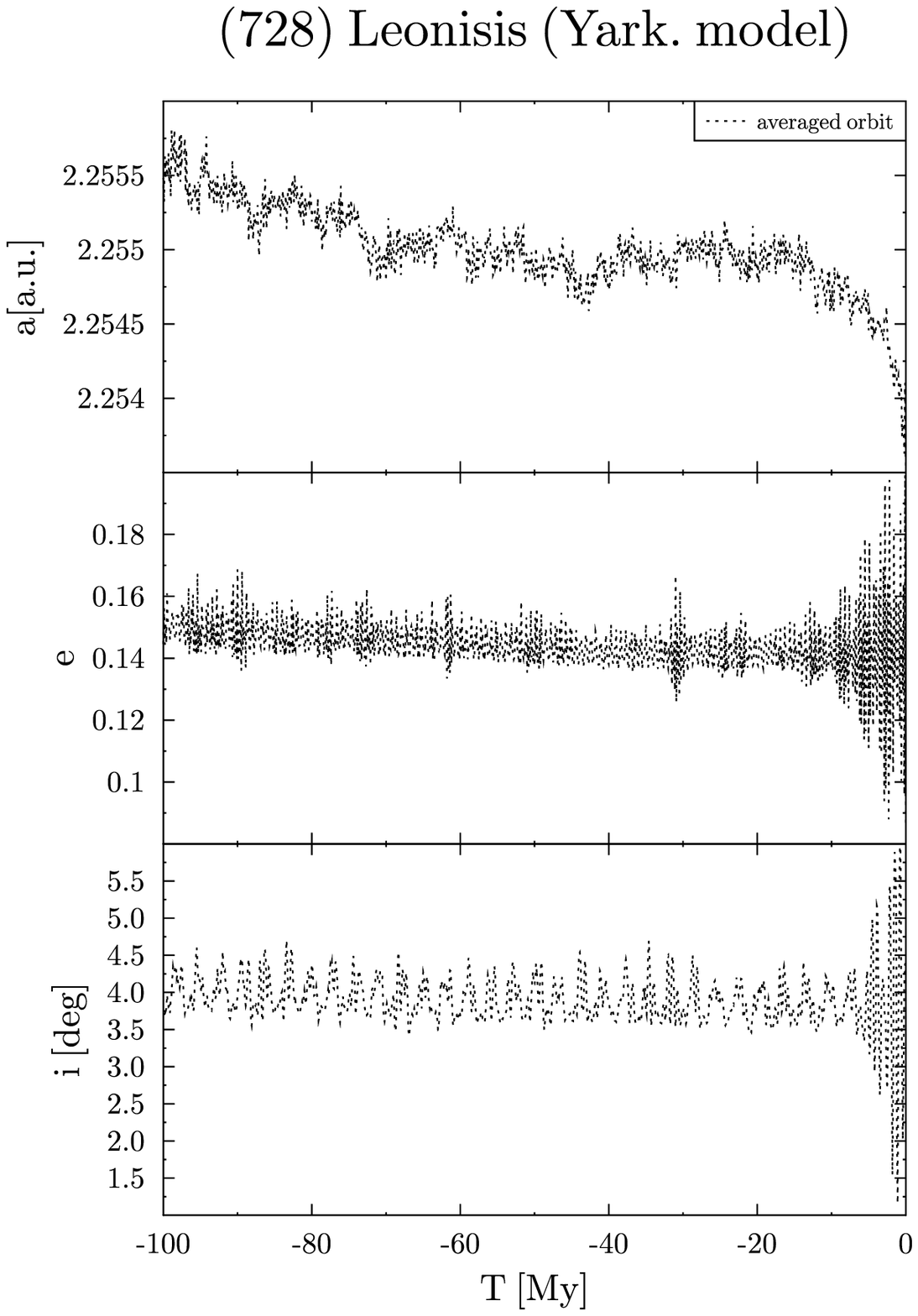}
\caption{Dynamic evolution of asteroid (728) Leonisis obtained with  the Yarkovsky model. \label{728evolution}}
\end{figure*}

\begin{figure*}[b]
\centering       
    \includegraphics[width=0.4\textwidth]{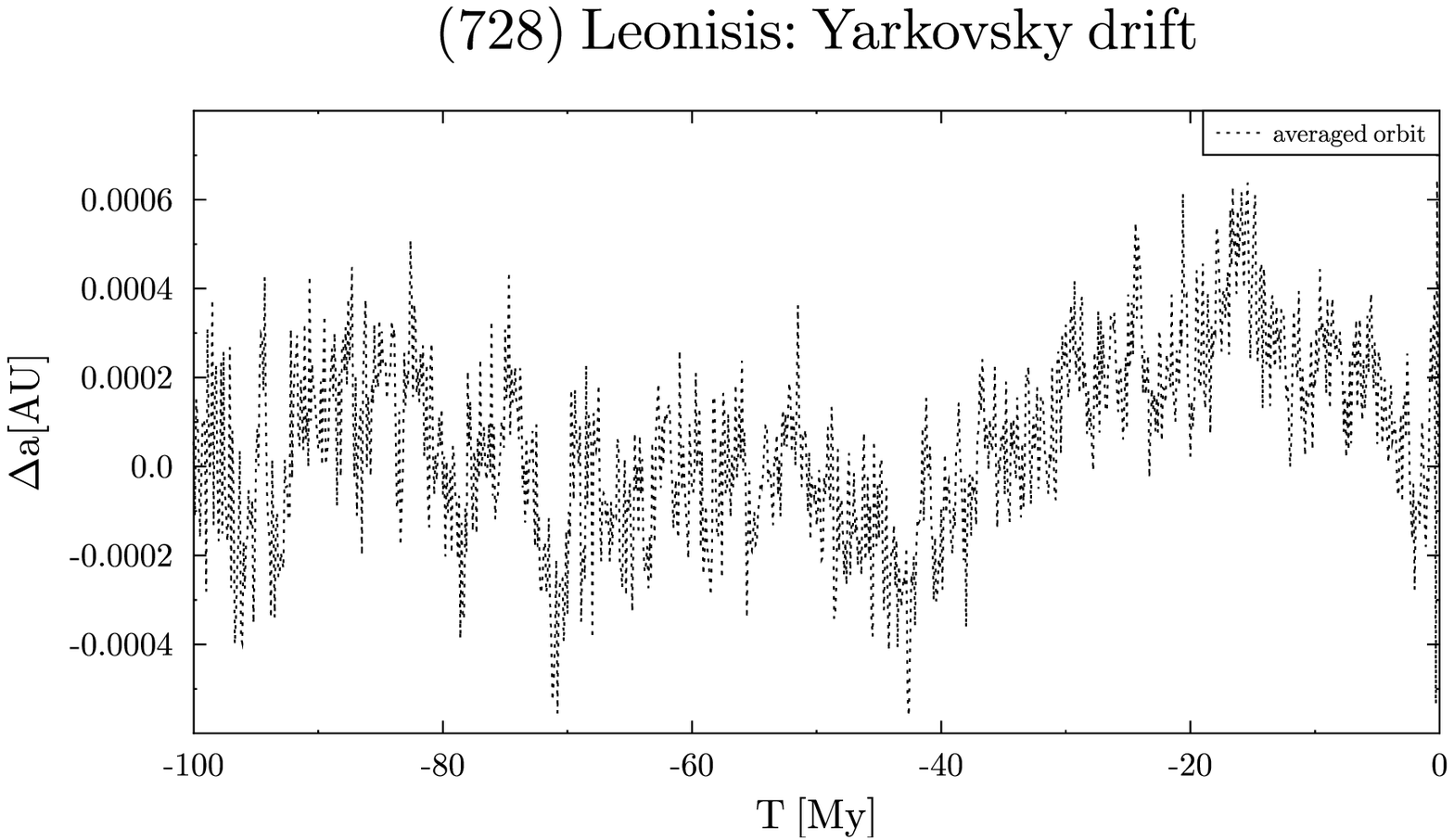}
    \caption{Drift in semi-major axis due to the Yarkovsky effect for asteroid (728) Leonisis. \label{728drift}} 
\end{figure*}

To picture the evolution of dynamical ties to the Flora and Vesta family in time, we calculated the distances from Vesta and Flora centroids ($D_{orbit}$, Parker et al. 2008) at each integration step. The centroids were defined based on the current family orbital element distributions. In Figs. \ref{122192}, \ref{lundia}, and \ref{leonis} we present the evolution of the average centroid distances during the past 100 My years for asteroids (122192), (809) Lundia, and (728) Leonisis. In Figs. \ref{122192evolution}, \ref{809evolution}, and \ref{728evolution} we plot the evolution of orbital elements, and in Figs. \ref{122192drift}, \ref{809drift}, and \ref{728drift} we show the Yarkovsky drift in the semi-major axis for the three objects. 

 Asteroid (122192) 2000 KK82 is a typical example of an object in the Flora region for which the dynamical history and centroid distances differ strongly depending on the selected model. The
orbit of (122192) is located close to the region of the 10:3 Jupiter resonance, but also to the three body 5J+4S-2A mean motion resonance (MMR) region. In the Yarkovsky model (Fig. \ref{122192d}),
the asteroid enters the resonant region about 15 My in the past due to the Yarkovsky drift in the semi-major axis. In the grav. model the asteroid never reaches that region. Moderate-order resonances play an important role in stabilizing this orbit. The mean Yarkovsky drift value is rather typical (average rate: 0.55e-5 AU/My) and changes during the integration. The asteroid shows smaller centroid distances to the Flora family than to Vesta family throughout the integration time. However, many of the physical parameters are unknown for (122192) 2000 KK82, and therefore a link to the Vesta family cannot be excluded for this object. 

Asteroid (809) Lundia is an example of an object where the differences in the evolutionary paths for the two models are small. In both dynamical models the asteroid reaches a stable orbit after around 15 My of integration into the past and remains stable throughout the integration after this. The Yarkovsky drift for (809) Lundia is the smallest among all the asteroids in our sample, and it remains relatively constant throughout the integration. According to \citet{carruba2005v}, Lundia is captured into a $(g-g_{6})+s-s_{6}$ secular resonance, also called the $z_2$ secular resonance. We note that Lundia is also very close to the 6:11 MMR with Mars. We compare our results for (809) Lundia with that of \citet{carruba2005v}. There are substantial differences in our approaches and results. \citet{carruba2005v} performed forward-integration staring with a position in the Vesta family and showed that migration to the current Lundia position is possible. In our approach we started with the current position of Lundia and performed back integration. To show the amplified drift effect, \cite{carruba2005v} created clones with either prograde rotation and $0^{\circ}$ obliquity or retrograde rotators with $180^{\circ}$ obliquity (for which the Yarkovsky drift is maximized) and size 100 m. In our modeling we used the observationally confirmed obliquity and spin axis of (809) Lundia, that is,  $\lambda=122^{\circ} \pm 5^{\circ} \; \beta=22^{\circ} \pm 5^{\circ}$ (\cite{Kryszczynska2014}) as well as the observationally constrained size (9.1 km) and albedo. Additionally, we used 100 clones as compared to two clones in \cite{carruba2005v}. We argue that using the observationally constrained sizes and spins gives us a better understanding of the actual dynamical history of this object. Based on the non-observationally confirmed assumptions, \cite{carruba2005v} suggested that (809) Lundia may have originated in the Vesta family and drifted to the Flora region. However, this scenario is only possible if retrograde spin clones are considered for (809) Lundia (see the two different paths in Fig. 7 in \cite{carruba2005v}). From observations we know that (809) Lundia has a prograde sense of rotation. Therefore out of the two paths shown in \cite{carruba2005v}, the prograde drifting toward the larger semi-major axis is more likely. Additionally, the size of Lundia (9.1 km) limits the amplitude of the Yarkovsky drift. For these two reasons, starting from a position in the Vesta family and slowly drifting toward the larger semi-major axis, Lundia would never have migrated to its current position in the Flora region.
This finding also agrees with \citep{nesvorny2008fugitives}, who showed that the Vesta fugitives in the inner main belt region should have preferentially retrograde rotations and thermal parameters that maximize the Yarkovsky drift amplitude. (809) Lundia with its prograde rotation and large size does not fit the picture of a typical fugitive.
Integrating back in time and taking the observational constraints into account, we find (809) Lundia on a stable orbit that is closer to the centroid of the Flora family than that of Vesta during the 100 My integration. Furthermore, (809) Lundia remains not only closer to the Flora centroid, but also outside the limit of the Vesta family throughout most of the integration time. Because of its prograde rotation, its relatively stable orbit during the 100 My, the small Yarkovsky drift in the semi-major axis, and its position on the V-shaped plot, we suggest that (809) Lundia is not a former member of the Vesta family.

This shows that the observational constraints and specifically the rotation direction need to be carefully considered when studying
the dynamical past and family membership of specific asteroids. Particularly the rotation direction has a direct implication on the direction of the Yarkovsky drift.

Another example of typical evolutionary path in the Flora region shows asteroid (728) Leonisis. The differences in the evolutionary paths for the two models are small. The average Yarkovsky drift is among the smallest in our sample and is variable though the integration time. Main, principal resonance in the vicinity is the 5:9 MMR with Mars (5M-9A, mentioned by Nesvorny et al., 2008a and Carruba, 2005). This is most probable cause of e and i variations in the last 10 My. The orbit of Leonisis is also close to 7:2 Jupiter mean-motion resonance. The object stays closer to the centroid of the Flora family throughout the entire integration than that of Vesta family. Unfortunately the Vesta family link cannot be excluded for this object due to unknown pole orientation, thermal and physical parameters.

For all the asteroids the main cause of the $e$,$i$ oscillations are multiple resonances, especially moderate-order two-body and three-body MMRs. The migration of orbital elements into resonant regions is often connected with the Yarkovsky drift. In the motion of (809) Lundia, the $z_2$ secular resonance plays an important
role. In the Yark. model all the considered asteroids have orbits that are closer to the centroid of the Flora family than to that of Vesta throughout the 100 My integration time. In Table \ref{parker} we list the centroid distances to the Flora and Vesta families for all the considered asteroids at T=-100 My and now (T=0 My). A more detailed analysis including more observational constrains needs to be made for all the objects. Observationally confirmed pole orientations in particular need to be considered. 

\section{Conclusions and future work}
\label{con}
The Flora family is one of the most taxonomically diverse asteroid families and has a rich dynamical history. Asteroids of various taxonomic types, albedos, color indices, and $G_{12}$s are present in this dynamical region, which suggests that they have various origins.
In particular, we here predicted 164 V-type asteroids in the dynamical region. These objects need to be targeted spectrally to confirm their taxonomical types. These asteroids in addition to spectrally confirmed V- and A-type objects in the Flora family may represent traces of a differentiated body other than asteroid (4) Vesta in the inner main belt region. To test this hypothesis, we investigated 11 asteroids from the region (of predicted or observationally confirmed V or A type) in more detail and showed that their distances to the centroid of the Flora family are smaller than those to the Vesta family during the last 100 My. An in-depth investigation is required to analyze the past of these objects beyond this timescale. 
However, many of the studied objects have very short Lapapunov times, leading to unfeasible orbits beyond the 100 My timescale and obscuring their dynamical past. For this reason, it is challenging or even impossible to reliably rule out or confirm their dynamical origin. Nonetheless, we suggest that some of these objects, such as (809) Lundia, are unlikely to originate from (4) Vesta. The dynamical investigation also showed that the Yarkovsky forces play a crucial role in the dynamical evolution and that the physical parameters have to be accurately known to better understand dynamical paths of those objects. Obtaining precise spin and shape models as well as determining thermal parameters is of great importance to constraining initial parameters for dynamical modeling and understanding evolutionary paths. In particular, we found that with the observational constraints, asteroid (809) Lundia, a V type object, probably does not originate in the Vesta family, as previously indicated by \cite{carruba2005v}. Because of its observationally constrained spin, the direction of the Yarkovsky drift, and its size, we suggest that the object did not migrate from the Vesta region. (809) Lundia does not fit the picture of a typical Vesta fugitive. We suggest that there might also be other objects similar to (809) Lundia in the inner main belt. In particular, V types with prograde rotation vs. mostly retrograte rotating fugitives need to be dynamically investigated. Future studies should investigate the dynamical history of (908) Lundia using a more complex binary model for the asteroid. The candidate asteroids should also be targeted spectrally to establish their spectral differences or similarities to the V types in the Vesta region and to the Bunburra Rockhole meteorite.


\begin{acknowledgements}
We would like to thank the anonymous referee for their highly appreciated comments and suggestions that significantly contributed to improving the quality of this publication.
DO was supported by the Polish National Science Centre, grant number 2012/04/S/ST9/00022. 
\end{acknowledgements}

\onecolumn

\begin{table}[htbp]
   \centering
   \begin{tabular}{|c|c|c|c|c|} \hline
        Group & Num. of obj. & $\overline{p_V}$ & $\overline{\sigma_{p_V}}$ & $std(p_V)$ \\ \hline  
Flora (all)                                                     & 3299 & 0.25 & 0.08 & 0.11 \\
Flora (without Baptistina)                              & 2472 & 0.27 & 0.08 & 0.11 \\
Flora (without Baptistina and clumps)   & 2191 & 0.27 & 0.08 & 0.11 \\
Baptistina                                              & 827 & 0.2 & 0.07 & 0.1 \\
Auravictrix                                             & 10 & 0.27 & 0.04 & 0.1 \\
Matterania                                              & 33 & 0.26 & 0.05 & 0.1 \\
Davidweilla                                             & 10 & 0.18 & 0.03 & 0.05 \\
Belgica                                                         & 18 & 0.33 & 0.07 & 0.11 \\
Albada                                                  & 12 & 0.24 & 0.06 & 0.09 \\
Hissao                                                  & 10 & 0.24 & 0.05 & 0.08 \\
Krinov                                                  & 11 & 0.31 & 0.11 & 0.06 \\
Amytis                                                  & 10 & 0.3 & 0.09 & 0.08 \\
Wolfernst                                                       & 10 & 0.29 & 0.06 & 0.11 \\ \hline
   \end{tabular}
   \caption{Median albedo in Flora clumps and subfamilies.}
   \label{averagepv}
\end{table}

\begin{table}[htbp]
   \centering
   \begin{tabular}{|c|c|c|c|c|c|c|c|} \hline
        Group & Num. of obj. & $\overline{a*}$ & $\overline{\sigma_{a*}}$ & $std_{a*}$ & $\overline{i-z}$ & $\overline{\sigma_{i-z}}$ & $std(i-z)$ \\ \hline    

Flora (all)                                                     & 4225 & 0.12 & 0.03 & 0.09 & -0.04 & 0.07 & 0.11 \\
Flora (without Baptistina)                              & 3389 & 0.13 & 0.03 & 0.09 & -0.05 & 0.07 & 0.11 \\
Flora (without Baptistina and clumps)   & 3143 & 0.13 & 0.03 & 0.09 & -0.05 & 0.07 & 0.11 \\

Baptistina                                              & 836 & 0.07 & 0.03 & 0.1 & -0.03 & 0.07 & 0.1 \\
Matterania                                              & 24 & 0.16 & 0.02 & 0.07 & -0.04 & 0.04 & 0.1 \\
Ingrid                                                  & 10 & 0.15 & 0.02 & 0.07 & -0.09 & 0.05 & 0.14 \\
Azalea                                                  & 23 & 0.16 & 0.02 & 0.03 & -0.04 & 0.04 & 0.06 \\
Lude                                                    & 10 & 0.13 & 0.02 & 0.09 & -0.03 & 0.04 & 0.08 \\
Alascattalo                                             & 19 & 0.13 & 0.02 & 0.1 & -0.06 & 0.04 & 0.09 \\
OAFA                                                    & 13 & 0.17 & 0.03 & 0.06 & -0.07 & 0.06 & 0.06 \\
Dornburg                                                        & 11 & 0.15 & 0.02 & 0.04 & -0.06 & 0.04 & 0.03 \\
Amytis                                                  & 17 & 0.16 & 0.02 & 0.05 & -0.34 & 0.05 & 0.19 \\\hline
  \end{tabular}
   \caption{Median a* in Flora clumps and subfamilies.}
   \label{averagea}
\end{table}

\begin{table}[htbp]
   \centering
   \begin{tabular}{|l|c|c|} \hline
        Designation             & Type                                          & Reference                                                                                               \\ \hline

        (298) Baptistina        & E, M, C, Xc                                   & \cite{florczak1998visible} \cite{mothe2005reanalysis}                            \\      
        (728) Leonisis                  & A / Ld                                                & \cite{florczak1998visible}                                                                      \\      (809) Lundia          & V                                                     & \cite{florczak2002discovering}                                                           \\
        (956) Elisa             & V                                                     & \cite{florczak2002discovering}                                                           \\
        (2780) Monnig &         A / Ld                                          & \cite{lazzaro2004s}                                                                             \\
        (2858) Carlosporter     & L                                                     & \cite{mothe2005reanalysis}                                                                       \\
        (3533) Toyota           & M, Xk                                         & \cite{florczak1998visible} \cite{xu1995small} or \cite{bus2002phase}          \\
        (3850) Peltier          & V                                                     & \cite{xu1995small} or \cite{bus2002phase}                                                \\
        (4278) Harvey           & V                                                     & \cite{florczak1998visible}                                                                       \\
        (4375) Kiyomori & A                                                     & \cite{mothe2005reanalysis}                                                                       \\
        (4556) Gumilyov         & A / L                                         & \cite{lazzaro2004s}                                                                             \\
        (4750) Mukai            & X                                                     & \cite{xu1995small} or \cite{bus2002phase}                                                \\
        (7255) 1993 VY1 & X                                                     & \cite{mothe2005reanalysis}                                                                       \\

  \hline        
   \end{tabular}
   \caption{Asteroids with spectra other than S  and C complex in the Flora family.}
   \label{other}
\end{table}

\twocolumn

\clearpage

\onecolumn
\begin{landscape}
\begin{table}[h]
\tiny
\caption{Keplerian elements (and rms uncertainties) of 11 selected asteroids; epoch: JD  2457000.5 TDT (2014-12-09).\label{table-elements}}
\label{orbital}
\begin{tabular}{|c|c|c|c|c|c|c|c|c|c|}
\hline
Ast.   name     &  Taxa &      $a[AU]$    &       $e$     &  $i_{2000}[deg]$ &     $\Omega_{2000}[deg]$    &       $\omega_{2000}[deg]$    &  $M[deg]$& no. of  & rms \\
  & & & & & & & & obs. used & [arc sec]  \\
\hline
(956) Elisa  & V [1]  & 2.29757& 0.205259& 5.96450& 192.664& 125.479& 290.205&   &        \\
1-$\sigma$ rms   &   & 8.02E-9& 5.04E-8& 5.72E-6& 5.01E-5& 5.18E-5& 1.56E-5& 1511& 5.24E-1 \\
(4278) Harvey & V [2] &  2.26683& 0.177584& 5.47574& 146.505& 235.233& 139.328&   &        \\ 
1-$\sigma$ rms    &  & 1.053E-8& 7.29E-8& 5.59E-6& 6.77E-5& 6.99E-5& 1.58E-5& 1119& 5.62E-1 \\  
(728) Leonisis  & A / Ld [3] &2.25391& 0.0876105& 4.25641& 82.6708& 55.1723& 142.944&   &        \\
1-$\sigma$ rms    &  & 6.79E-9& 3.93E-8& 5.28E-6& 7.17E-5& 7.96E-5& 3.54E-5& 1364& 4.98E-1 \\  
(4375) Kiyomori & A [3] [4]  &2.29351& 0.104495& 5.86327& 137.557& 69.8264& 311.666&   &        \\ 
1-$\sigma$ rms      && 1.35E-8& 8.57E-8& 6.29E-6& 7.66E-5& 8.27E-5& 3.78E-5& 1033& 5.25E-1 \\  
(2780) Monnig & A / Ld [3]  &2.19419& 0.115843& 5.46732& 274.991& 312.466& 100.482&   &        \\
1-$\sigma$ rms   &   & 9.85E-9& 4.91E-8& 5.67E-6& 5.93E-5& 6.56E-5& 2.87E-5& 1067& 5.14E-1 \\
(4556) Gumilyov & A / L[3] &2.31040& 0.145536& 4.75199& 111.901& 213.755& 290.406&   &        \\
1-$\sigma$ rms      & & 1.44E-8& 5.60E-8& 5.34E-6& 7.34E-5& 7.70E-5& 2.87E-5& 1169& 5.45E-1 \\
(809) Lundia  & V [1] & 2.28258& 0.193118& 7.14897& 154.583& 196.249& 265.245&   & \\
1-$\sigma$ rms      &    & 4.86E-9& 5.09E-8& 4.71E-6& 4.24E-5& 4.48E-5& 1.64E-5& 1719& 5.09E-1 \\
(122192) 2000 KK82 & V* & 2.33976 &0.146215& 5.73099& 199.474& 207.051& 218.271&   &        \\
1-$\sigma$ rms      & & 2.04E-8& 1.06E-7& 1.30E-5& 6.54E-5& 7.29E-5& 3.42E-5& 347& 6.13E-1 \\
(124360) 2001 QR133& V* & 2.32323& 0.183149& 5.21932& 248.974& 68.8619& 284.170&   &        \\
1-$\sigma$ rms      & &2.08E-8& 7.71E-8& 7.56E-6& 8.25E-5& 8.82E-5& 3.49E-5& 406& 5.69E-1 \\
(151563) 2002 TM135 & V* & 2.22461& 0.177543& 6.96351& 122.275& 256.778& 243.590&   &        \\
1-$\sigma$ rms   &   & 1.56E-8& 9.10E-8& 6.77E-6& 7.63E-5& 7.98E-5& 2.48E-5& 210& 5.69E-1 \\
(103418) 2000 AM150 & V* & 2.26591& 0.140663& 6.94947& 140.812& 272.373& 185.026&   &        \\ 
1-$\sigma$ rms  &    & 1.94E-8& 6.74E-8& 9.52E-6& 6.12E-5& 7.29E-5& 4.25E-5& 276& 5.27E-1 \\ 
\hline
\multicolumn{10}{|l|}{[1] \cite{florczak2002discovering} }\\
\multicolumn{10}{|l|}{[2] \cite{florczak1998visible}  }\\
\multicolumn{10}{|l|}{[3] \cite{lazzaro2004s}} \\
\multicolumn{10}{|l|}{[4] \cite{mothe2005reanalysis}    }\\
\multicolumn{10}{|l|}{*predicted, this work}\\
\hline
\end{tabular}
\end{table}

\begin{table}[h]
\tiny
\caption{Thermal properties of 11 selected asteroids.\label{table-thermal}}
\begin{tabular}{|c|c|c|c|c|c|c|c|c|c|}
\hline
 Ast. name            &     $r$    &        $\rho_{bulk}$  &    $\rho_{surf.}$  &  $\kappa$  & $C$  &  $p_v$ &   $\epsilon$  & P & spin \\
            &   $[m]$   &        $[kg/m^3]$  &    $[kg/m^3]$  &  $[W/K/m]$ & $[J/kg/K]$ &  $(pV)$ &     &   [h] &  \\

\hline
(956) Elisa      &   5237 $\pm$ 800       &   3000 [7]          &        1500 [8]       &           0.001 [8]         &              680 [9]       &                 0.1468 $\pm$ 0.0221 [5]     &   0.95   [8]      &           16.492 $\pm$ 0.001 [6]        &      random  \\
(4278) Harvey    &   2245 $\pm$ 227       &   3000 [7]         &        1500  [8]      &           0.001  [8]        &               680 [9]       &                 0.318 $\pm$ 0.035[1]   &  0.95   [8]      &           3.118 $\pm$ 0.315 [2]     &      random  \\
(728) Leonisis   &   3634 $\pm$ 328       &   2700 [7]         &        1500  [8]      &           0.001  [8]        &               680 [9]       &                 0.2557 $\pm$ 0.0272 [5]     &  0.95   [8]      &           5.5783 $\pm 0.0002$ [6]        &      random  \\
(4375) Kiyomori  &   2920 $\pm$ 835       &   2700 [7]         &        1500  [8]      &           0.001  [8]        &               680 [9]       &                 0.298 $\pm$ 0.143 [1]   &  0.95   [8]      &           6.4709 $\pm$ 0.0004 [6]        &      random  \\
(2780) Monnig    &   3045 $\pm$ 253       &   2700 [7]         &        1500  [8]      &           0.001  [8]        &               680 [9]       &                 0.2287 $\pm$ 0.0206 [5]     &  0.95   [8]      &           4.229 $\pm$ 0.350 [2]      &     random \\
(4556) Gumilyov  &   2612 $\pm$ 257       &   2700 [7]          &       1500  [8]      &           0.001  [8]        &               680 [9]       &                 0.2825 $\pm$ 0.0321[5]      & 0.95   [8]      &           3.628 $\pm$ 0.356 [2]     &      random \\
(809) Lundia     &   4550 $\pm$ 500 [3]     &   2500 $\pm$ 200 [4]       &       1500  [8]       &           0.001 [8]          &               680  [9]       &                 0.3266 $\pm$ 0.0572 [5]   &
0.95    [8]     &           15.41574 $\pm 0.00001$ [4]  &      $\lambda=122^{\circ} \pm 5^{\circ} \; \beta=22^{\circ} \pm 5^{\circ}$[4] \\
(122192) 2000 KK82      &      748   $\pm$  76    &    3000 [7]          &       1500  [8]      &           0.001  [8]       &                680   [9]      &                 0.318 $\pm$ 0.035[1]   &  0.95    [8]        &        1.0388 $\pm$ 0.105 [2]     &     random \\
(124360) 2001 QR133      &      807  $\pm$  82     &    3000 [7]          &       1500 [8]       &           0.001 [8]        &                680  [9]      &                 0.318 $\pm$ 0.035[1]   &  0.95   [8]         &        1.1201 $\pm$ 0.113 [2]     &     random \\
(151563) 2002 TM135      &      523  $\pm$  53     &    3000 [7]          &       1500 [8]       &           0.001 [8]        &                680  [9]      &                 0.318 $\pm$ 0.035[1]   &  0.95  [8]          &        0.7262 $\pm$ 0.073 [2]     &     random \\
(103418) 2000 AM150      &      630  $\pm$  64     &    3000 [7]           &      1500 [8]       &           0.001 [8]        &                680  [9]      &                 0.318 $\pm$ 0.035[1]    & 0.95   [8]         &        0.8749 $\pm$ 0.088 [2]     &     random \\
\hline
\multicolumn{10}{|l|}{$r[m]$ - if not determined: calculated by diameter formula by \cite{Fowler1992}}\\
\multicolumn{10}{|l|}{[1] Typical (mean) values of albedos determined by \cite{Usui2013}}\\
\multicolumn{10}{|l|}{[2] rotation period, if not determined: $P[sec] = 5(D[m]/2) = 5 R [m]$.}\\
\multicolumn{10}{|l|}{[3] Equivalent of $r$ value for Lundia (single-body approximation).}\\
\multicolumn{10}{|l|}{[4] \cite{Kryszczynska2014}}\\
\multicolumn{10}{|l|}{[5] \cite{masiero2011main}}\\
\multicolumn{10}{|l|}{[6] JPL Small body browser.}\\
\multicolumn{10}{|l|}{[7] Typical densities of V and A asteroids, used as default \citep{Carruba2014}}\\
\multicolumn{10}{|l|}{[8] Commonly used surface densities, $\epsilon$ and $\kappa$ values \citep{Carruba2014}, \citep{Broz2006}}\\                                                                                    
\multicolumn{10}{|l|}{[9] Values expected for regolith-covered MBA. \citep{Broz2006}.}\\
\multicolumn{10}{|l|}{We adapted densities, $\kappa$, $C,$ and $\epsilon$ values used by most authors without uncertainties (as typical, default values).}\\
\hline
\end{tabular}
\end{table}

\end{landscape}

\begin{table}
\caption{Average values of the Yarkovsky drift rates in the semi-major axes of 11 selected asteroids.}

\begin{center}
\begin{tabular}{|c|c|}
\hline
 Name            &     $da/dt$ $[AU/My]$   \\

\hline
956 Elisa      &   0.20e-05  \\
4278 Harvey    &   0.27e-05  \\
728 Leonisis   &   0.32e-05  \\
4375 Kiyomori  &   0.78e-05  \\
2780 Monnig    &   0.80e-05 \\
4556 Gumilyov  &   0.37e-05 \\
809 Lundia     &   0.17e-05 \\   
(122192)      &    0.55e-05   \\
(124360)      &    0.86e-05    \\
(151563)      &    0.29e-05   \\
(103418)      &    0.43e-05   \\
\hline
\end{tabular}
\label{drifts}
\end{center}
\end{table}

\begin{table}
\caption{Vesta and Flora $D_{orbit}$ criteria for different models, present and past ($T=-100$ My) values. Vesta boundary according to \cite{Parker2008} (The maximum deviation in orbital space, in units of adopted dispersions, all values rounded to two decimal places): $D_{orb, V}=2.75$; Flora boundary: $D_{orb, F}=2.6$.}

\begin{center}
\begin{tabular}{|c|c|c|c|}
\hline
\multicolumn{4}{|c|}
{$D_{orb}$ for Vesta family} \\
\hline
 Name            &   $D_{orb} (T=0)$  &  $D_{orb} (T=-100My)$ & $D_{orb} (T=-100My)$ \\
                      &                              & (grav. model)                   & (Yark. model) \\
\hline
956 Elisa               & 7.11  & 5.13 & 4.78 \\
4278 Harvey             & 5.67  & 4.27 & 4.37 \\
728 Leonisis            & 4.70  & 6.61 & 6.31 \\
4375 Kiyomori   & 1.54  & 4.89 & 4.39 \\
2780 Monnig             & 3.20  & 3.77 & 4.72 \\
4556 Gumilyov   & 4.61  & 4.23 & 4.91 \\
809 Lundia              & 6.39  & 3.94 & 3.57 \\ 
(122192)                & 3.45  & 4.83 & 2.41 \\
(124360)                & 6.07  & 6.25 & 3.24 \\
(151563)                & 5.51  & 3.84 & 3.63 \\
(103418)                & 3.06  & 4.40 & 3.62 \\
\hline
\multicolumn{4}{|c|}
{$D_{orb}$ for Flora family} \\
\hline
 Name            &   $D_{orb} (T=0)$  &  $D_{orb} (T=-100My)$ & $D_{orb} (T=-100My)$ \\
                      &                              & (grav. model)                   & (Yark. model) \\ \hline
956 Elisa               & 2.82  & 2.07 & 1.90 \\
4278 Harvey             & 1.71  & 0.99 & 0.96 \\
728 Leonisis            & 1.78  & 1.15 & 0.95 \\
4375 Kiyomori   & 1.40  & 1.87 & 1.10 \\
2780 Monnig             & 1.44  & 1.57 & 1.64 \\
4556 Gumilyov   & 0.73  & 1.18 & 1.13 \\
809 Lundia              & 2.84  & 1.64 & 1.58 \\   
(122192)                & 1.35  & 2.31 & 1.46  \\
(124360)                & 2.02  & 2.81 & 1.48 \\
(151563)                & 2.42  & 1.69 &  1.68 \\
(103418)                & 1.63  & 1.97 &  1.82 \\
\hline

\end{tabular}
\label{parker}
\end{center}
\end{table}

\twocolumn

\newpage
\bibliography{biblio}
\bibliographystyle{aa}

\end{document}